\documentclass[twocolumn,showpacs,prl]{revtex4}

\usepackage{bm}

\usepackage{graphicx}

\usepackage{subfigure}

\begin{document}

\title{Signatures of Strong Momentum Localization via Translational-Internal Entanglement}

\author{Nir Bar-Gill and Gershon Kurizki}

\affiliation{Weizmann Institute of Science, 76100
Rehovot, Israel}

\date{\today }


\begin{abstract}
We show that atoms or molecules subject to fields that couple their internal and translational (momentum) states may undergo a crossover from randomization (diffusion) to strong localization (sharpening) of their momentum distribution. The predicted crossover should be manifest by a drastic change of the interference pattern as a function of the coupling fields.
\end{abstract}

\maketitle


It is well known that potential energy disorder added to a spatial lattice of energy degenerate sites may cause localization of an otherwise delocalized (hopping) quantum mechanical particle \cite{Anderson}. This localization is crucially dependent not only on the amount of disorder, but also on the lattice dimensionality \cite{Anderson,Akkermans,Akulin}. Momentum localization may be caused by time-dependent perturbations \cite{fishman}. Here, we consider the hitherto unexplored analogs of these phenomena in the {\em momentum space of a diffracted particle}, due to {\em correlations of internal and translational states} within the particle. We show that such translational-internal entanglement (TIE) may incur disorder that causes a crossover from diffusion to strong localization of the momentum distribution and thereby a drastic change in the diffraction pattern, even for particles with few internal levels.

The first point we must address is: how to create the intra-particle momentum-space analog of a perfectly ordered lattice, in which all sites have equal energy? The free particle energy-momentum relation implies that each momentum state carries a different energy. An ordered lattice can however be achieved, by correlating (entangling) a selected set of {\em discrete} momentum states with eigenstates of different degrees of freedom. This means that the corresponding states $| \vec{k}_n, n \rangle$, where $\hbar \vec{k}_n$ stands for the momentum of state $n$, and $\varepsilon_n$ for the energy eigenvalues of its additional degrees of freedom, satisfy in an ordered, momentum-space ``lattice''
\begin{equation}
\frac{\hbar^2 k_n^2}{2M} + \varepsilon_n = \frac{\hbar^2 k_{n'}^2}{2M} + \varepsilon_{n'} \label{en_eq}
\end{equation}
for all $n,n'$. 
It will be shown later that such TIE may also yield momentum lattices which deviate from condition (\ref{en_eq}) and have {\em controllable dimensionality and disorder}. Alternatively to TIE, one may entangle orthogonal momentum components of the particle. However, this has the limitation of complicating the resulting diffraction pattern (see below).

\begin{figure}
\subfigure[]{
\includegraphics[width=0.9\linewidth,height=0.5\linewidth]{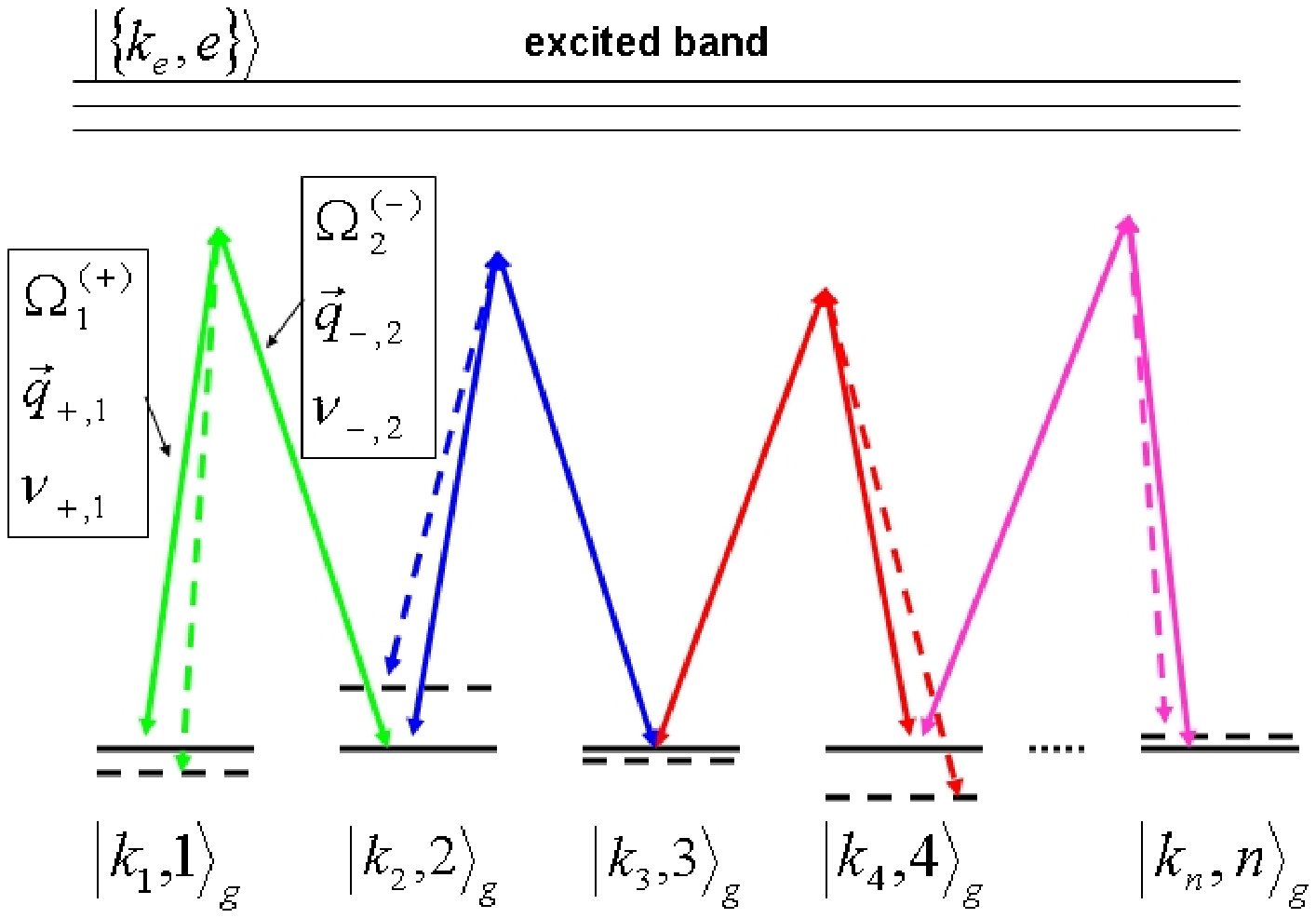}}
\\
\subfigure[]{
\includegraphics[width=0.9\linewidth,height=0.3\linewidth]{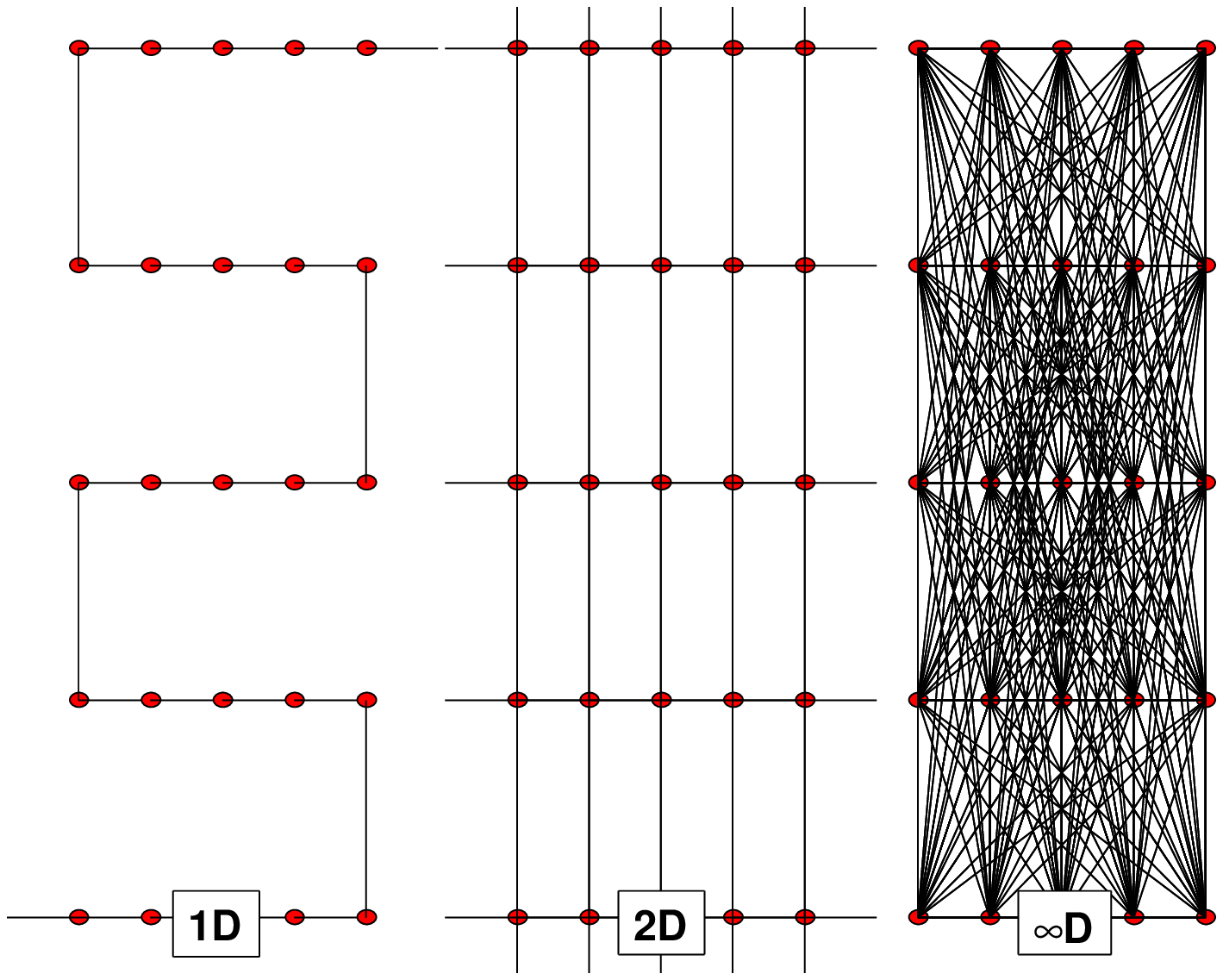}}
\subfigure[]{
\includegraphics[width=0.9\linewidth,height=0.5\linewidth]{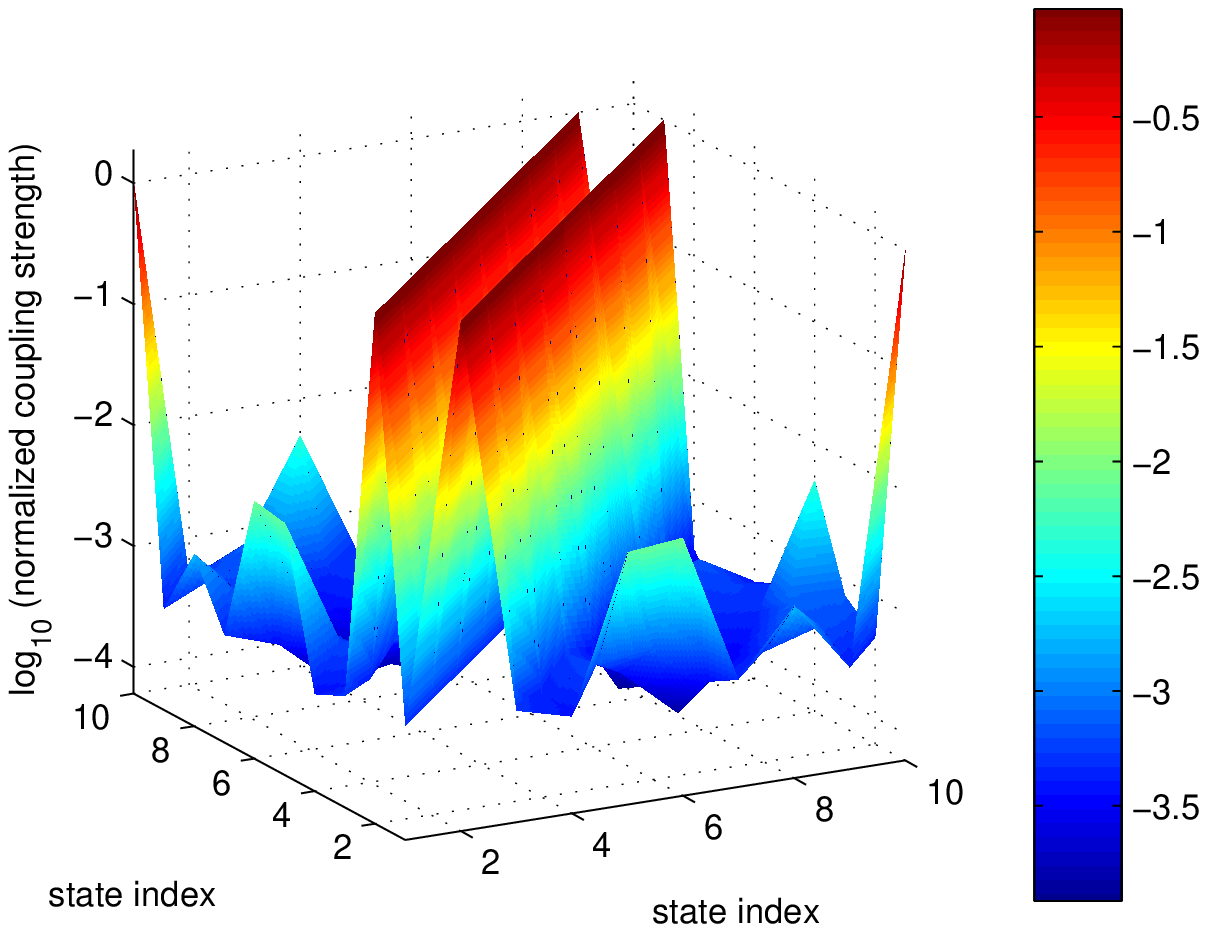}}
\caption{(a) Proposed setup: the internal states of a cold atom (or molecule) are mixed by pairs of Raman beams and become entangled with different momentum states. Full lines indicate an ordered lattice of energies, dashed lines - a disordered lattice. (b) Momentum-space dimensionalities $1D, 2D, \infty D$, illustrated by lattice graphs with different connectivities. (c) Numerical Raman coupling strengths in log scale for Li atoms in a $1D$ setup (for parameters at paper end).}
\label{setup}
\end{figure}

In order to create such a TIE particle, consider a cold atom or molecule (assuming a non-interacting ensemble thereof) initially prepared in a single internal state $| 0 \rangle$ and in a narrow wavepacket of momentum states centered around $\hbar \vec{k}_0$.
The state $|\vec{k}_0,0 \rangle$ can next be coupled to a band of $N$ long-lived, non-degenerate states $| \vec{k}_n,n \rangle$ via Raman near-resonant pairs of laser beams (see Fig. \ref{setup}(a)), analogously to \cite{chu}. In turn, these $N$ states may be quasi-adiabatically Raman-coupled among themselves. Examples of appropriate systems are atomic ground-state hyperfine/Zeeman sublevels, multiplets of circular Rydberg states, or rovibrational bands of a molecular ground state. Following the outlined quasi-adiabatic Raman sequence, the system occupies a ``lattice'' of entangled momentum-internal states, whose dimensionality and disorder are controllable by the Raman couplings, as detailed below.

The system Hamiltonian $H_S$ and the dipolar system-field interaction Hamiltonian $H_I$ in the Rotating Wave Approximation (RWA) are \cite{Scully}
\begin{eqnarray}
H_S &=& \hbar \sum_{n=1}^N \omega_n^{(g)} |\vec{k}_n,n \rangle_g \langle \vec{k}_n,n |_g \nonumber \\
H_I &=& \hbar \sum_{n=1}^N \left[ \Omega_{ne}^{(+)} e^{-i (\nu_{+,n} t + \vec{q}_{+,n} \cdot \vec{r} )} | \left \{ \vec{k}_e,e \right \} \rangle \langle \vec{k}_n,n |_g \right. \\
&+& \left. \sum_{n'=1}^N \Omega_{en'}^{(-)} e^{i (\nu_{-,n'} t + \vec{q}_{-,n'} \cdot \vec{r})} | \vec{k}_{n'},n' \rangle_g \langle \left \{ \vec{k}_e,e \right \} | \right] + h.c. \nonumber
\end{eqnarray}
Here, $|\vec{k}_n,n \rangle_g$ are the states of the stable (long-lived) band with momenta $(\hbar \vec{k}_n)_g$ and energies $\hbar \omega_n^{(g)}$, whereas $| \left \{ \vec{k}_e,e \right \} \rangle$ refer to states of a far-detuned, {\em intermediate} unstable (electronically-excited) band. The $+$ and $-$ steps of the Raman coupling consist respectively of virtual (off-resonant) transitions $|\vec{k}_n,n \rangle_g \leftrightarrow |\left \{ \vec{k}_e,e \right \} \rangle \leftrightarrow |\vec{k}_n',n' \rangle_g$ via pairs of laser beams with frequencies $\nu_{+,n}$ and $\nu_{-,n'}$ and respective wavevectors $\vec{q}_{+,n}, \vec{q}_{-,n'}$. The corresponding Rabi frequencies are $\Omega_{ne}^{(+)}$ and $\Omega_{en'}^{(-)}$. The frequencies are chosen such that $\nu_{+,n} - \nu_{-,n'} \simeq \omega_{n'}^{(g)} - \omega_n^{(g)}$. Hence the two-step Raman process $|\vec{k}_n,n \rangle_g \rightarrow |\vec{k}_{n'},n' \rangle_g$ is near-resonant only for a chosen pair of laser beams, causing energy and momentum transfer of $\hbar (\nu_{+,n} - \nu_{-,n'})$ and $\hbar (\vec{q}_{+,n} - \vec{q}_{-,n'})$, respectively. In the frame transforming away the laser frequencies and wavevectors and upon {\em adiabatically eliminating} \cite{Scully} the intermediate unstable states $| \left \{ \vec{k}_e,e \right \} \rangle$ from the Schroedinger equation yields the following pairwise coupling equations for the eigenstate amplitudes $c_n$ and $c_{n'}$
\begin{eqnarray}
\dot{c}_{n'} & \simeq & \frac{i}{\hbar} (E_{n'} c_{n'} + J_{nn'} c_n), \nonumber \\
E_{n'} &=& \hbar (\nu_{+,n} - \nu_{-,n'} + \omega_n^{(g)} - \omega_{n'}^{(g)}), \nonumber \\
J_{nn'} & \simeq & \hbar (\Omega_{ne}^{(+)} \Omega_{en'}^{(-)})/\left[ \nu_{+,n} - (\omega^{(e)} - \omega_n^{(g)}) \right].
\label{H3}
\end{eqnarray}
Here we have assumed sufficiently weak fields to neglect power broadening corrections (AC Stark shifts), as well as the off-resonant linewidths of the far-detuned fields $\gamma_{eg} \left[ \Omega_{n,e}^{(\pm)}/(\nu_{\pm,n} - \omega^{(e)} - \omega_{n}^{(g)}) \right]^2 << \gamma_{eg}$.

Even with the near-resonant Raman selectivity imposed on (\ref{H3}), it allows for rich, complex dynamics. Here we wish to map (\ref{H3}) onto known models of strong localization \cite{Anderson}. To this end, we require:
\begin{equation}
\nu_{+,n} - \nu_{-,n'} + \omega_n^{(g)} - \omega_{n'}^{(g)} = const. ; J_{nn'} = J = const.
\label{H4}
\end{equation}
These requirements amount to adjusting the frequencies $\nu_{\pm,n}$ and the two-step Rabi-frequency product to be independent of $n,n'$. The momenta $\hbar \vec{k}_{n'} = \hbar (\vec{k}_n + \vec{q}_{+,n} - \vec{q}_{-,n'})$ are separately controlled to give equal diagonal energies (ordered ``lattice'') or random energies $E_n$ (disorderd ``lattice''). The disorder is measured in terms of the width $\Delta$ of the flat ({\em uncorrelated}) distribution of on-diagonal energies, such that the random $E_n \in [-\Delta/2,\Delta/2]$. 

Our setup allows the creation of momentum-space configurations with {\em any effective dimensionality}. This dimensionality is not related to the spatial dimensionality of the atomic or molecular ensemble, but rather to the off-diagonal coupling terms. 
A 1D momentum-space lattice is created by resonantly coupling states $| \vec{k}_n,n \rangle$ with the states $| \vec{k}_{n+1},n+1 \rangle$ and vice-versa. In order to fulfill periodic boundary conditions, in a system of finite size $N$, we couple $| \vec{k}_N,N \rangle \leftrightarrow | \vec{k}_1,1 \rangle$. The single-particle Hamiltonian describing the system in the entangled basis $|\vec{k}_i,i \rangle$ is represented by the matrix
\begin{eqnarray}
H = \left (
\begin{array}{cccccc}
	E_1 & -J & 0 & 0 & \cdots & -J \\
	-J & E_2 & -J & 0 & 0 & \cdots \\
	0 & -J & E_3 & -J & 0 & \cdots \\
	\vdots & & & & & \\
	-J & 0 & \cdots & 0 & -J & E_N 
\end{array}
\right). \label{H1d}
\end{eqnarray}
The ability to satisfy Eqs. (\ref{H3}), (\ref{H4}) so that (\ref{H1d}) is realized is numerically demonstrated in Fig. 1(c): laser beam pairs selectively couple level pairs with equal $J$ but random $E_n$.

By adding Raman resonant laser-beam pairs, such that each state (``lattice site'') is coupled to an increasing number of other ``sites'', the Hamiltonian emulates a system of higher dimensionality. Thus, 1D, 2D and 3D ``lattices'' are realized when each ``site'' has 2, 4 and 6 neighbors, respectively (see Fig. \ref{setup}(b)). Specifically, we assume the number of sites $N$ equals the system size $L$ in 1D, $L^2$ in 2D and $L^3$ in 3D. Therefore, the neighbors of site $i$ in 1D are $i \pm 1$, in 2D the neighbors $i \pm L$ are added, and in 3D - $i \pm L^2$ are added. This dimensionality argument holds as long as all $N$ ``sites'' have the same connectivity \cite{Anderson,Akkermans,Akulin}, as we assume in the following.
In the $N \rightarrow \infty$ (thermodynamic) limit, this system is equivalent to either a perfect or a disordered infinite lattice, depending on $\Delta$.
\begin{figure}
\includegraphics[width=0.8\linewidth,height=0.4\linewidth]{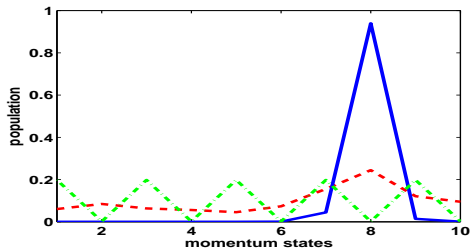}
\caption{Momentum distribution for random systems with $(\Delta/J)=1$ (red-dashed) and $(\Delta/J)=10$ (blue-solid), in a 1D system with $N=10$ states, compared to a periodic (regular) system with $(\Delta/J)=10$ (green-dotted). The random momentum distribution is delocalized for $(\Delta/J)=1$, while strong localization appears for $(\Delta/J)=10$. The periodic distribution is not localized for $(\Delta/J)=10$. A localization ``crossover'' is found even though a random 1D system should not exhibit a transition, due to finite size effects.}
\label{mom_dist_1d}
\end{figure}

As is well known, in 1D any amount of random disorder causes localization \cite{Anderson,Akkermans,Akulin}.
In Fig. \ref{mom_dist_1d} we present the results of our 1D lattice calculations for the adiabatic ramping-up of the two-photon coupling $J$, and different values of the diagonal disorder $\Delta$. We plot the ground-state 1D momentum distribution for $N=10$ different states $|\vec{k}_n,n \rangle$, a setup not too difficult to realize. It can be seen that finite size effects, resulting from the finite number of states N, are manifest as an {\em artificial} ``localization crossover'' in this 1D configuration (this occurs also in 2D): the localization length increases with a reduction in the strength of the disorder, until, for small enough disorder, the localization length exceeds the length of the system, causing the momentum distribution to appear delocalized. 

\begin{figure}
\subfigure[]{
\includegraphics[width=0.48\linewidth,height=0.4\linewidth]{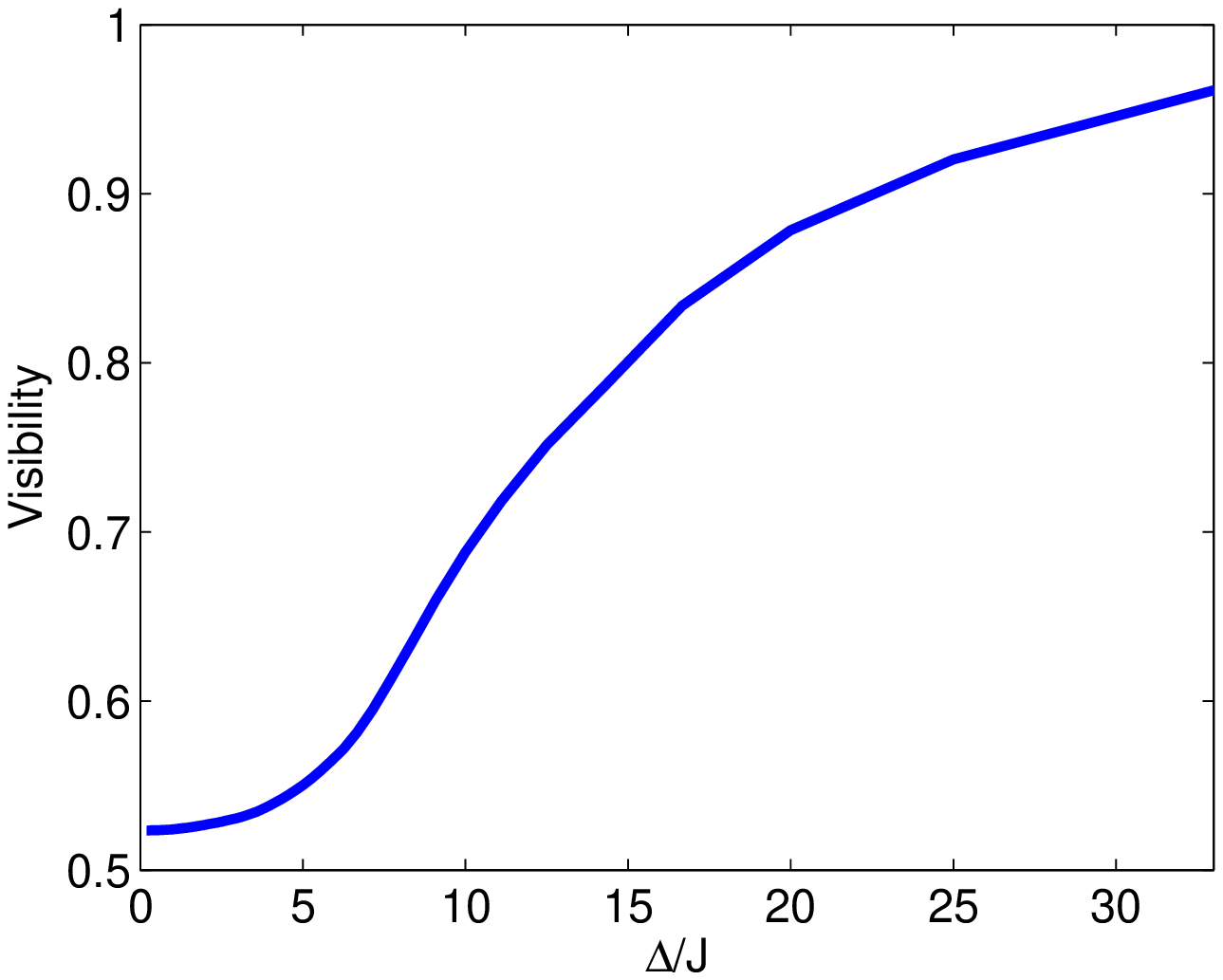}}
\subfigure[]{
\includegraphics[width=0.48\linewidth,height=0.4\linewidth]{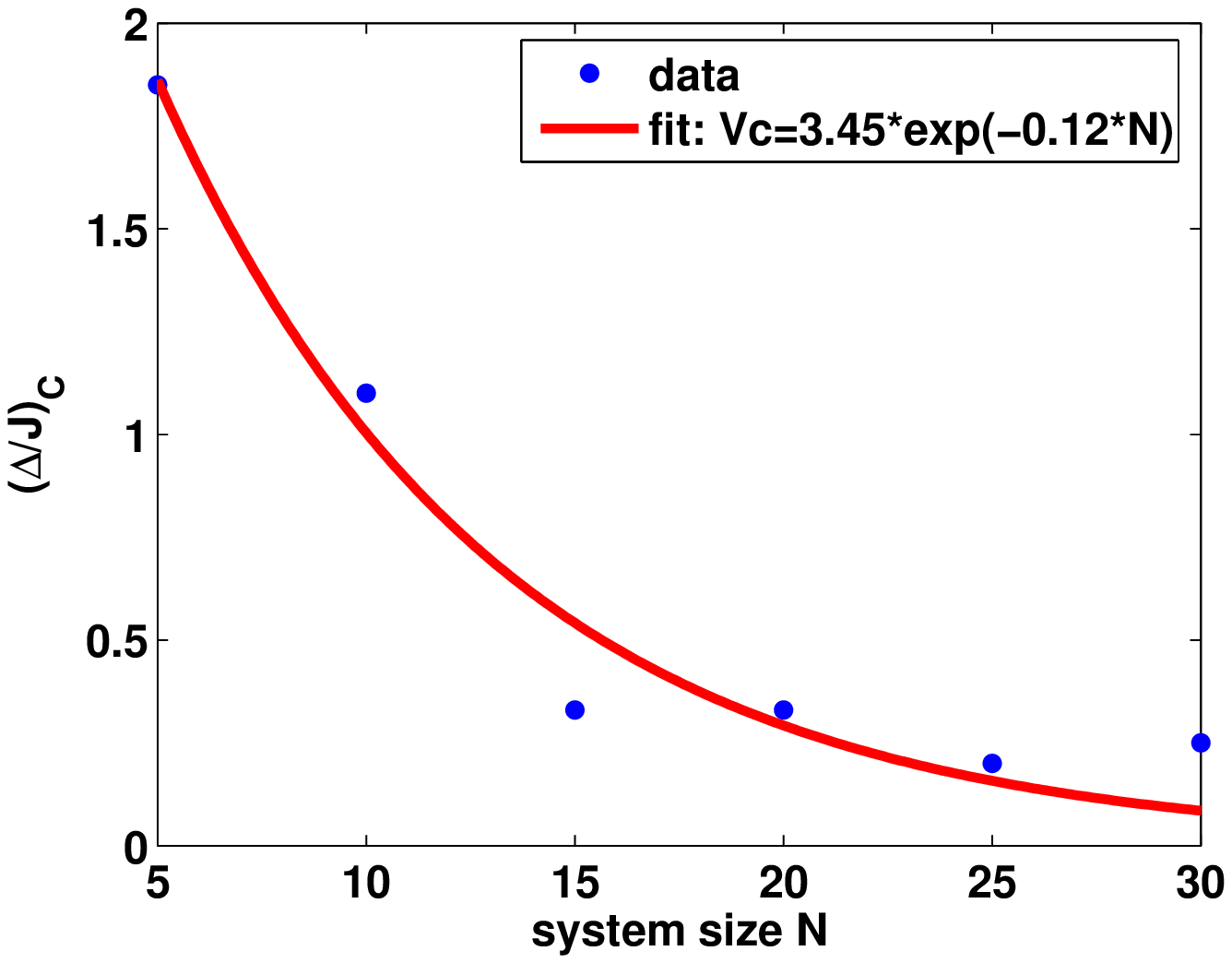}}
\subfigure[]{
\includegraphics[width=0.48\linewidth,height=0.4\linewidth]{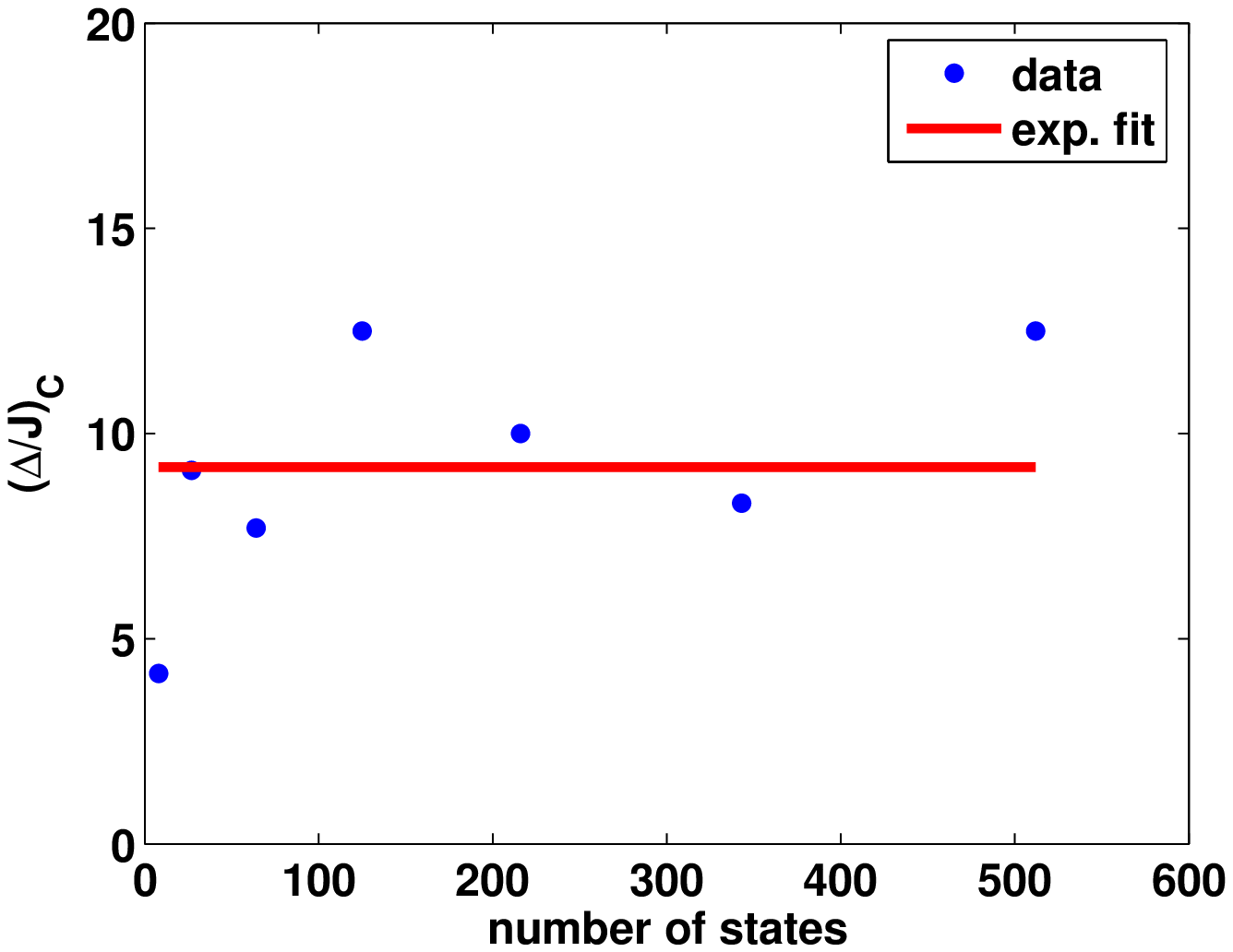}}
\subfigure[]{
\includegraphics[width=0.48\linewidth,height=0.4\linewidth]{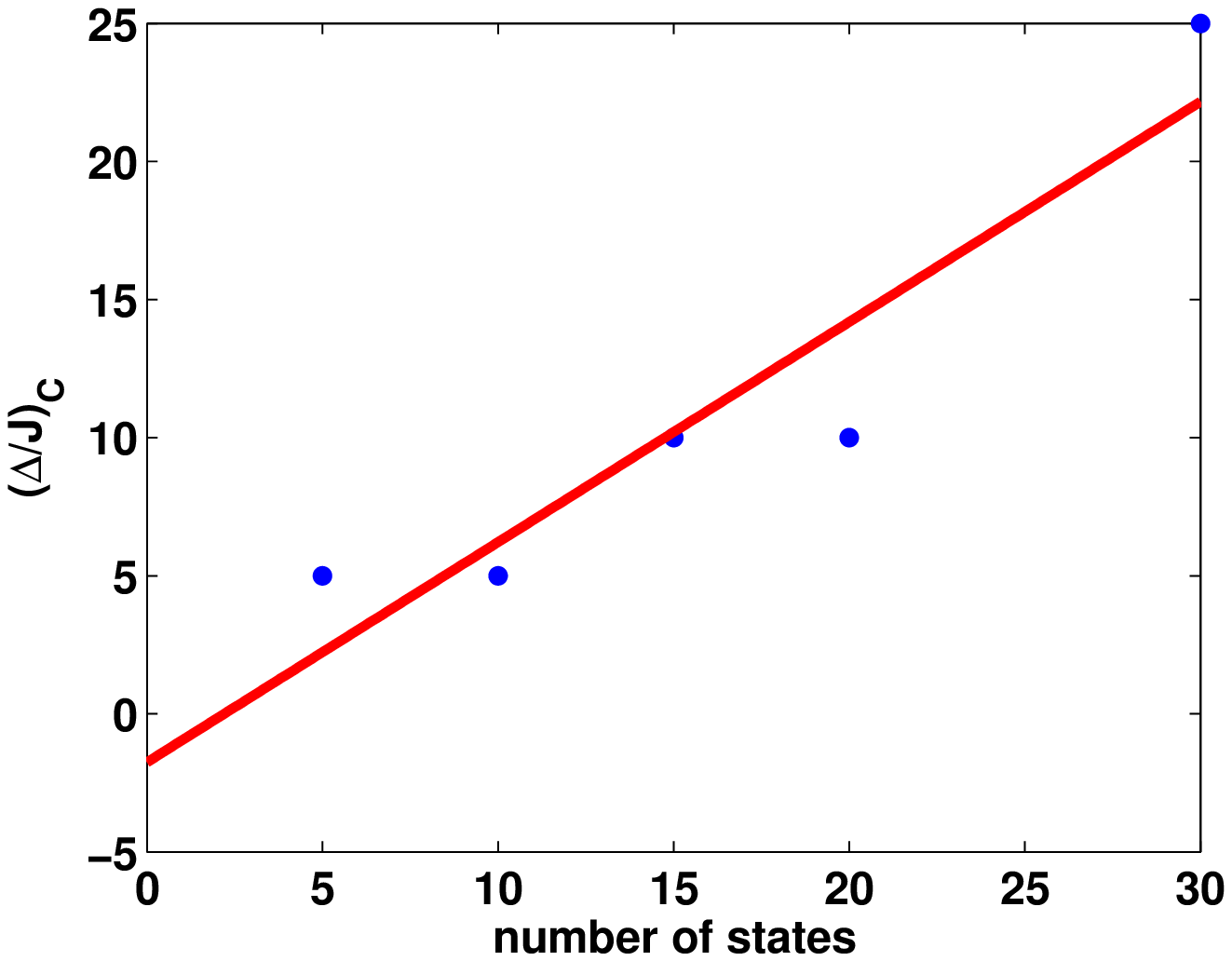}}
\caption{(a) Visibility of the interference fringes as a function of $\Delta/J$, in a 3D system. (b) Flow of the ``crossover'' point to zero in a 1D system, calculated (blue dots) and exponentially fitted (red line). (c) Same, for a 3D system. The ``crossover'' point flows to $(\Delta/J)_C \simeq 10$, indicating a true transition. (d) Same for an $\infty D$ system: ``crossover'' flows to $\infty$.}
\label{flow_1d}
\end{figure}

In order to quantify the localization crossover for each effective dimensionality, one can use the ground-state momentum distribution, and apply such standard measures as the entropy or the Inverse Participation Ratio (IPR) \cite{Scalettar}. 
We suggest a different localization measure, appropriate for atomic or molecular interferometry \cite{Berman,Arndt}:
(a) Consider an internally structured particle that is incident upon an interferometer in a state with a narrow momentum distribution $f (\vec{k} - \vec{k_0})$ around $\vec{k_0}$ and an arbitrary superposition of internal-energy eigenstates $| n \rangle$:
\begin{eqnarray}
| \psi_{in} \rangle &=& \int d \vec{k} f (\vec{k}-\vec{k_0}) | \vec{k} \rangle \sum_n c_n | n \rangle; \nonumber \\
\langle \vec{r} | \psi_{in} \rangle &\simeq& e^{i \vec{k_0} \cdot \vec{r}} \sum_n c_n | n \rangle. \label{in}
\end{eqnarray}
In a Mach-Zehnder Interferometer (MZI), it is split between two alternative, interfering paths, $\vec{r}_1$ and $\vec{r}_2$. Because of the internal-translational {\em factorization}, such a state can exhibit a high-visibility interference pattern \cite{Arndt}, since its propagation in the MZI results in
\begin{equation}
\langle \vec{r} | \hat{U_0} |\psi_{in} \rangle = e^{i \vec{k_0} \cdot (\vec{r_1} - \vec{r_2})} \sum_n c_n |n\rangle. \label{standard}
\end{equation}
(b) By contrast, the output state resulting from TIE propagation in the MZI is
\begin{equation}
\langle \vec{r} | \psi_{out} \rangle = \langle \vec{r} | \hat{U}_{TIE} |\psi_{in}\rangle = \sum_n c_n e^{i \vec{k_n} \cdot ( \vec{r_1} - \vec{r_2} ) } | n \rangle, \label{tie_out}
\end{equation}
where $c_n$ and $\vec{k}_n$ are subject to Eqs. (\ref{H3})-(\ref{H4}) above.
The averaging of the detection probability of state (\ref{tie_out}) over $|n\rangle$ tends to wash out the interference fringes:
\begin{eqnarray}
Tr_n \{ \langle \vec{r} | \psi_{out} \rangle \langle \psi_{out} | \vec{r} \rangle \} = \sum_n |c_n |^2 cos^2 [\vec{k}_n \cdot (\vec{r}_1 - \vec{r}_2)]. \label{vis}
\end{eqnarray}
Thus, the width of the momentum distribution of such a TIE particle is directly related to the visibility of the interference fringes measured by passing this particle through a MZI. Specifically, for a flat distribution of $|c_n|^2$, the interference pattern (\ref{vis}) is $\frac{1}{2} \left[ 1+ sinc(k_0 L) \right]$, assuming $k_n = k_0 n$ and system size $L$. Thus, the visibility scales as $1/L$, approaching zero for $L \rightarrow \infty$. By contrast, for a localized distribution $|c_n|^2 \sim e^{-\gamma n}$, the interference pattern becomes $\frac{1}{2} + \frac{2 \gamma}{4 \gamma^2 + 4 k_0^2} \left[ \gamma cos \left(k_0 L \right) + k_0 sin \left( k_0 L \right) \right]$. The visibility is then $\frac{4 \gamma (\gamma + k_0)}{4 \gamma^2 + 4 k_0^2}$, i.e. it does not depend on the system size $L$ and it approaches $1$ in the localized limit $\gamma >> k_0$. In Fig. \ref{flow_1d}(a) we plot the visibility of the interference pattern as a function of the disorder $\Delta/J$ for a 3D system, showing a crossover from a delocalized state (low visibility) to a localized state (high visibility).

It is possible to distinguish between an artificial transition caused by finite size effects and a true transition, by checking the {\em scaling} of the ``crossover'' point $(\Delta/J)_C$ as a function of system size. The ``crossover'' point is found upon adiabatically ramping up the coupling for different values of $\Delta/J$, and calculating the visibility for the resulting momentum distribution. A true transition should occur at the discontinuity point of the derivative of the visibility, i.e. it should jump from zero to its maximal value. Due to finite-size effects, the smoothed ``crossover'' is at the point of the {\em maximal derivative} of the visibility (as a function of $\Delta/J$), which remains continuous. 

In Fig. \ref{flow_1d}(b) it can be seen that for 1D the ``crossover'' point flows toward $(\Delta/J)_C=0$ (this occurs for 2D as well), indicating that there is no thermodynamical transition and the system is always localized.
A 3D momentum-space lattice, which is more difficult to realize experimentally due to the large number of Raman pairs needed ($N=L^3$, where $L$ is the system size in each dimension), exhibits a true localization transition as $N \rightarrow \infty$. In Fig. \ref{flow_1d}(c) we plot the transition point as a function of $N$, displaying the finite size scaling. In this case the transition point flows toward a {\em finite, non-zero value}.

Can we emulate an infinite-dimensional (but finite-size) momentum-space lattice? Instead of the matrix (\ref{H1d}), infinite dimensionality corresponds to non-zero coupling terms in all off-diagonal elements. Such a setup would obviously require a prohibitively large number of Raman pairs. However, by coupling every level to the initially populated level, for which the number of Raman pairs needed is equivalent to that of the 1D case, the system is governed by the Hamiltonian
\begin{eqnarray}
H = \left (
\begin{array}{cccccc}
	E_1 & -J & -J & -J & \cdots & -J \\
	-J & E_2 & 0 & 0 & 0 & \cdots \\
	-J & 0 & E_3 & 0 & 0 & \cdots \\
	\vdots & & & & & \\
	-J & 0 & \cdots & 0 & 0 & E_N 
\end{array}
\right). \label{Hnd}
\end{eqnarray}
which is similar to an infinite dimensional system of states $n=2...N$ (in the sense that all eigenstates are thermodynamically delocalized). As in the 1D case, it is known that the Anderson model of infinite dimensionality does not have a phase transition \cite{Ulmke}. However, here the system will remain delocalized, with the ``crossover'' point flowing to infinity (or, more precisely, scaling as $N$). This is shown in Fig. \ref{flow_1d}(d), where we plot the transition point as a function of the system size.


An experimental demonstration may involve ultracold Li atoms, which allow significant momentum to be imparted by laser beams. The atoms can be outcoupled from a Bose-Einstein condensate, and prepared in an initial state $|F=1,m_F=-1,k_0 \rangle$, with a velocity of $v_0 \simeq 10 \frac{cm}{sec}$ and an energy of $E_0 \simeq 74 kHz$. This state can then be Raman-coupled to the Zeeman-split $m$ states of levels $F=1$ and $F=2$, providing a total of $8$ accessible levels. Such a {\em small number of levels} cannot reproduce 3D effects, but {\em can provide measurable scaling results} for the 1D Hamiltonian (\ref{H1d}) and the effective $\infty$D Hamiltonian (\ref{Hnd}). These levels are accessible using pairs of laser beams (far detuned by hundreds of GHz from the single-photon resonance), and detuned from each other with an accuracy of $\sim 1kHz$ by means of acousto-optic modulators (AOMs). These laser beam pairs have been numerically shown to create momentum states with energies in the range of $0 - 300 kHz$, whose separation allows the neglect of off-resonant couplings (unaccounted for by Eq. (\ref{H3}) - see Fig. 1c). Random energies are realized by randomly setting the angles between the beams. The desired coupling strengths can be achieved with beam power of a few mW and beam waists of $\sim 100 \mu m$. The AOMs allow quasi-adiabatic control over the coupling Raman beams, as required in order to probe ground state properties. A Mach-Zehnder interferometer can be realized by two perpendicular pairs of standing-wave laser beams, forming the two beam splitters of the interferometer (as in \cite{rempe}). The interference fringes are then recorded by counting the number of atoms in the two scattered clouds in a time-of-flight image. Molecular or Rydberg atom experiments with larger $N$ may be feasible \cite{Arndt}, but merit separate discussion of conditions (\ref{H3}) and (\ref{H4}).

To conclude, we have studied an intriguing fundamental effect: strong localization of the momentum distribution of particles subject to TIE and inter-state mixing. It may be revealed by interferometry of such particles. Remarkably, even few-level diffracted particles allow for measurable scaling effects that bear the signature of strong localization.

We acknowledge the support of the EC (QUACS RTN and SCALA NOE) and ISF.




\begin{thebibliography}{99}

\bibitem{Anderson} P. W. Anderson, Phys. Rev. \textbf{109}, 5, 1492 (1958).

\bibitem{Akkermans} G. Montambaux and E. Akkermans, {\it Physique Mesoscopique des Electrons et des Photons}, EDP Sciences (2004).

\bibitem{Akulin} V. M. Akulin, {\it Coherent Dynamics of Complex Quantum Systems}, Springer (2006).

\bibitem{fishman} P. J. Bardroff et. al, Phys. Rev. Lett. \textbf{74}, 3959 (1995); F. L. Moore, J. C. Robinson, C. Bharucha, P. E. Williams and M. G. Raizen, Phys. Rev. Lett. \textbf{73}, 2974 (1994); S. Fishman, D. R. Grempel and R. E. Prange, Phys. Rev. Lett. \textbf{49}, 509 (1982).

\bibitem{chu} M. Kasevich and S. Chu, Phys. Rev. Lett. \textbf{67}, 181 (1991).

\bibitem{Scully} M. O. Scully and M. S. Zubairy, {\it Quantum Optics} (Cambridge
University Press, 1997); C. Cohen-Tannoudji, {\it Atomic Motion in Laser Light}, (Les Houches, 1990).

\bibitem{Scalettar} R. T. Scalettar, G. G. Batrouni and G. T. Zimanyi, Phys. Rev. Lett \textbf{66}, 24, 3144 (1991).

\bibitem{Arndt} V. Buzek, M. Hillery and L. Mlodinov, Phys. Rev. A \textbf{71}, 062104 (2005); M. Arndt et al., Nature \textbf{401}, 680 (1999).



\bibitem{Berman} P. Berman Ed., Atom Interferometry, Academic Press, NY (1997).

\bibitem{Ulmke} M. Ulmke, V. Janis and D. Vollhardt, Phys. Rev. B \textbf{51}, 16, 10411 (1995).

\bibitem{rempe} S. D\"{u}rr, T. Nonn, and G. Rempe, Nature
\textbf{395}, 33 (1998).


\end{thebibliography}
\end{document}